\documentclass[a4paper,12pt]{article}
\usepackage{amssymb}
\usepackage{amsthm}
\usepackage{latexsym}
\usepackage{amsmath}
\usepackage{amssymb}
\usepackage{indentfirst}
\usepackage{graphicx}

\newcommand{\udots}{\mathinner{\mskip1mu\raise1pt\vbox{\kern7pt\hbox{.}}
		\mskip2mu\raise4pt\hbox{.}\mskip2mu\raise7pt\hbox{.}\mskip1mu}}
\usepackage{amsmath}
\newtheorem{definition}{Definition}

\newtheorem{lemma}{Lemma}[section]
\newtheorem{remark}{Remark}[section]
\newtheorem{proposition}{Proposition}[section]

\begin{document}
\title{\bf \Large {\bf    Best- and worst-case Scenarios for  GlueVaR distortion risk measure with Incomplete information    }}
{{\author{\normalsize{Mengshuo  Zhao}\\{\normalsize\it    (School of Statistics and Data Science, Qufu Normal University}\\
			\noindent{\normalsize\it Shandong 273165, China)}\\
			\normalsize{Chuancun Yin}
			\thanks{Corresponding author.}\\
			{\normalsize\it  (School of Statistics and Data Science,  Qufu Normal University}\\
			\noindent{\normalsize\it Shandong 273165, China}\\
			email:  ccyin@qfnu.edu.cn)
		}
		\date{}
		\maketitle
		
		\noindent{\large {\bf Abstract}} This paper derives the best- and worst-case GlueVaR distortion risk measure within a unified framework, based on partial information of the underlying distributions and shape information such as symmetry. In addition, we characterize the extremal distributions of GlueVaR with convex envelopes of the corresponding distortion functions. As examples, extremal cases of VaR, TVaR and RVaR are derived.
		
		

		\baselineskip =17pt

		
		\numberwithin{equation}{section}
		\section{Introduction}\label{intro}
		\noindent
		The problem of deriving the sharp bounds for risk measures under partial information of the underlying distributions is not new and has been subject to numerous articles, which we typically refer to as extremal cases risk measure problem. When the first two moments of the underlying distributions are known, closed-form solutions for worst-case VaR and worst-case TVaR were obtained, respectively (see, e.g., EI Ghaoui et al.
		(2003) and Chen et al. (2011)). In a broader scope, also under the constraints of the first two moments,  Li (2018) has done pioneering work to derive worst-case closed-form solutions for a broad class
		of risk measures, known as law-invariant coherent risk measures. Liu et al. (2020) determined the worst-case values of a law-invariant convex risk functional when
		the mean and a higher order moment such as the variance of a risk are known. Cai et al. (2023) generalized the results of Li (2018) and Liu et al. (2020) to
		the case of any distortion risk measure. Das  et al. (2021) characterized properties of the optimal order quantities in a newsvendor model within a robust framework that accounts for distributional ambiguity with first and $\alpha \text{th}$ moment constraints. Beyond the moments information, Popescu (2005) showed how to effectively calculate the optimal moment bounds for convex distributions that satisfy special properties such as symmetry and unimodality. Li et al.
		(2018) derived the sharp upper bounds of the worst-case RVaR with the first two moments and symmetry. Zhu and Shao (2018) further generalized the result of Li et al. (2018)
		to both worst-case and best-case bounds of distortion risk measures. Shao and Zhang (2023a,
		2023b) utilizd the first two moments (or certain higher-order absolute
		center moments) and the symmetry of underlying distributions to derive closed-form solutions for distortion risk measures in extreme
		cases. Furthermore, they characterized the distributions of the corresponding extremal cases using the envelopes of the distortion functions. Bernard et al. (2023) derived quasi explicit best- and worst-case values of a large class of distortion risk measures
		when the underlying loss distributions has a given mean and variance and lies within a  $\sqrt{\varepsilon}$-Wasserstein ball around a given reference distribution, which generalizes the results of   Li (2018) and Zhu and Shao (2018) corresponding to a Wasserstein tolerance of $\varepsilon=+\infty$. Cai et al. (2024a) investigated the worst-case values of the distortion
		risk measures of the stop-loss $(X-d)_+$ and limited loss $X\wedge d$ when
		the distribution of an underlying loss random variable $X$ is uncertain, see also Cai et al. (2024b) for same lines of this topic. Recently, Zhao et al. (2024) obtained closed-form solutions for the extreme
		distortion risk measures, both worst-case and best-case, based on only the
		first two moments and shape information such as the symmetry/unimodality
		property of the underlying distributions.
	
	The GlueVaR family was first introduced and studied by Belles-Sampera et al.(2014), which
	forms part of a wider class referred to as distortion
	risk measures. The well known VaR, TVaR and RVaR can be seen as special cases of GlueVaR distortion risk measure. Moreover, $\text{GlueVaR}^{h1,h2}_{\beta,\alpha}(X)$ can be expressed as a linear combination of
	$\text{TVaR}_{\beta}(X)$, $\text{TVaR}_{\alpha}(X)$, and $\text{VaR}_{\alpha}(X)$. This paper studies best- and worst-case bounds for GlueVaR distortion risk measure and their corresponding attainable distributions. Under our results, the extreme-case VaR, extreme-case TVaR and extreme-case RVaR can be easily derived. Our contribution can be summarized as follows. Firstly, we obtain the closed-form solutions for the best- and worst-case GlueVaR distortion risk measure and the corresponding extreme-case distributions by the knowledge of the first two moments of the underlying distributions. Secondly, in addition to the first two moments, considering the symmetry of the underlying distributions, we can still obtain the extremal cases estimates of GlueVaR distortion risk measure. Our findings broaden the applicability of closed-form solutions for extremal cases of distortion risk measures, and significantly improve existing moment bounds of the GlueVaR distortion risk measure under both moment and shape constraints, thus enriching the GlueVaR knowledge system.
		
		The subsequent sections of this paper are structured as follows. The necessary notations  are briefly described in Section 2 and the main problems are formulated. In Section 3, the analytic worst-estimated expressions and their corresponding distributions for GlueVaR distortion risk measure under the first two moments and symmetry properties are calculated,
		from which special cases are obtained: worst-case VaR, worst-case TVaR and worst-case RVaR, and the best estimation expressions and their corresponding distributions are given in Section 4. Finally, Section 5 draws conclusions and outlines future work.
		\section{ Preliminaries}\label{intro}
		Assume all random variables in this paper are defined on a
		common probability space $L^2(\Omega,\mathcal{F},\mathbb{P})$. The cumulative distribution function and the decumulative distribution function of a r.v. X are denoted by $F_X$  and $\bar{F}_X$, respectively. We use the notation $F_X^{-1}(p)$ to denote the left-continuous generalized inverse of $F_X$, given by $F_X^{-1}(p)=\inf\{x:F_X(x)\ge p\}, 0< p\le 1$. Meanwhile, we denote by $F_X^{-1+}(p)=\sup\{x:F_X(x)\le p\}, 0\le p<1$, the right-continuous generalized inverse of $F_X$. The left- and right-continuous inverse distribution function only possibly differ at countably many points in $[0,1]$.
		\begin{definition}[Distortion Risk Measure]
		Consider a random variable $X$ with $F_X=P(X\le x)$,
		the distortion risk measure (DRM) of $X$  is defined via
		the Choquet integral in the form
		$$\rho_h[X]=\int_0^{\infty}h(\bar{F}_X(x))dx+\int_{-\infty}^0 (h(\bar{F}_X(x))-1)dx,$$
		whenever at least one of the two integrals is finite. The function $h$ refers to a distortion function which is a nondecreasing function on $[0,1]$ with $h(0)=0$ and $h(1)=1$.	
		\end{definition}
		For convenience, we define $\tilde{h}(p)=1-h(1-p)$ for all $p\in [0,1]$ for the distortion function $h$, which is clearly also a distortion function, and call it the dual distortion function.
		\begin{definition}[GlueVaR Distortion Risk Measure (Belles-Sampera et al. (2014))]
		$\rho_h$ is called the GlueVaR distortion risk measure if the distortion function $h$ follows
		\begin{eqnarray}
			\mathcal{K}_{\beta,\alpha}^{h_1,h_2}(p)=\left\{\begin{array}{ll}
				\frac{h_{1}p}{1-\beta}, \ &{\rm if}\, \,  p\in [0,1-\beta),\\
				h_1+\frac{(h_2-h_1)[p-(1-\beta)]}{\beta-\alpha},  \ &{\rm if}\,\, p\in [1-\beta,1-\alpha),\\
				1, \ &{\rm if}\, \,  p\in [1-\alpha,1].\\
			\end{array}
			\right.
		\end{eqnarray}
		where the constants $\alpha, \beta, h_1$ and $h_2$ satisfy $0<\alpha<\beta<1$ and $0\le h_1\le h_2<1$. The shape of the  GlueVaR distortion function is determined by the distorted survival probabilities $h_1$ and $h_2$ at levels $1-\beta$ and $1-\alpha$, respectively.
		\end{definition}
		GlueVaR distortion risk measure can be expressed as a linear combination of three risk measures: TVaR at confidence levels $\beta$ and $\alpha$ and VaR at confidence
		level $\alpha$ (details can be found in Belles-Sampera et al. (2014)):
		$$\text{GlueVaR}^{h1,h2}_{\beta,\alpha}(X)= {\omega}_1 \times \text{TVaR}_{\beta}(X)
		+ {\omega}_2 \times \text{TVaR}_{\alpha}(X)
		+ {\omega}_3 \times \text{VaR}_{\alpha}(X),$$
where $\omega_1=h_1-\frac{(h_2-h_1)\times (1-\beta)}{\beta-\alpha}$, $\omega_2=\frac{h_2-h_1}{\beta-\alpha}\times (1-\alpha)$, $\omega_3=1-\omega_1-\omega_2=1-h_2$.

		Note that $\text{VaR}_\alpha$, $\text{TVaR}_\alpha$ and $\text{RVaR}_{\alpha,\beta}$ are particular cases of the GlueVaR risk measure. In this framework, by calibrating the parameters on the heights $h_1$ and $h_2$ and the parameters $\alpha$ and $\beta$, the distortion functions for these classic risk measures are obtained as follows. For a random variable $X$, let $h_1=h_2=0$, $\beta=\alpha$ ($\alpha\in(0,1)$), the  distortion function for VaR is
		\begin{eqnarray*}
			\mathcal{K}_{\alpha,\alpha}^{0,0}(p)=\left\{\begin{array}{ll}
				0,  \ &{\rm if}\,\, p\in [0,1-\alpha),\\
				1, \ &{\rm if}\, \,  p\in [1-\alpha,1].\\
			\end{array}
			\right.
		\end{eqnarray*}
	Let $h_1=h_2=1$,$ \beta=\alpha$ ($\alpha\in(0,1)$), the distortion function for TVaR is
\begin{eqnarray*}
	\mathcal{K}_{\alpha,\alpha}^{1,1}(p)=\left\{\begin{array}{ll}
		\frac{p}{1-\alpha}, \ &{\rm if}\, \,  p\in [0,1-\alpha),\\
		1, \ &{\rm if}\, \,  p\in [1-\alpha,1].\\
	\end{array}
	\right.
\end{eqnarray*}
Let $h_1=0,h_2=1$, the distortion function for RVaR is
\begin{eqnarray*}
	\mathcal{K}_{\beta,\alpha}^{0,1}(p)=\left\{\begin{array}{ll}
		0, \                                 & {\rm if}\, \,  p\in [0,1-\beta),      \\
		\frac{p-(1-\beta)}{\beta-\alpha},  \ & {\rm if}\,\, p\in [1-\beta,1-\alpha), \\
		1, \                                 & {\rm if}\, \,  p\in [1-\alpha,1].
	\end{array}
	\right.
\end{eqnarray*}
Obviously, $\lim\limits_{\beta\uparrow 1}\mathcal{K}_{\beta,\alpha}^{0,1}(p)=\mathcal{K}_{\alpha,\alpha}^{1,1}(p)$, $\lim\limits_{\beta\downarrow \alpha}\mathcal{K}_{\beta,\alpha}^{0,1}(p)=\mathcal{K}_{\alpha,\alpha}^{0,0}(p)$.\\


Figure 2 illustrates $\widetilde{\mathcal{K}_{\beta,\alpha}^{h_1,h_2}}(p)$, which is the dual distortion function of GlueVaR distortion risk measure. In this figure, given a piecewise dual distortion function $\widetilde{\mathcal{K}_{\beta,\alpha}^{h_1,h_2}}(p)$ that consists of three distinct segments,  it is straightforward to compute the slope of its second segment as $\frac{h_2-h_1}{\beta-\alpha}$ and the slope of its third segment as $\frac{h_1}{1-\beta} $. For ease of reference, we let \(k_1=\frac{h_2-h_1}{\beta-\alpha}\), \(k_2=\frac{h_1}{1-\beta}\).

 Figure 3 illustrates $\mathcal{K}_{\beta,\alpha}^{h_1,h_2}(p)$, the distortion function of GlueVaR distortion risk measure, which is a piecewise linear distortion function with three segments. It is intuitive that the slope of the first segment is \(k_2=\frac{h_1}{1-\beta}\) and the slope of the second segment is \(k_1=\frac{h_2-h_1}{\beta-\alpha}\). For convenience of exposition, we let $k_3=\frac{1-h_2}{\alpha}$, where $k_3$ is the slope of the straight line $\vec{AB}$.

\begin{definition}[Symmetry]
A random variable X (or its corresponding distribution) is called $\mathit{symmetric}$ if there is a constant $m$ such that $\mathbb{P}(X\le x)=\mathbb{P}(X\ge 2m-x)$ for any $x\in \Bbb{R}$, and we call $m$ the $\mathit{symmetric}$ $\mathit{center}$.
\end{definition}
\begin{definition}[Boyd and Vandenberghe (2004)]
 For a distortion function $\mathcal{K}$, the convex and concave envelopes of $\mathcal{K}$ are defined, respectively, by
$$\mathcal{K}_*=\sup\{g| g:[0,1]\rightarrow [0,1] \, {\rm  is\,\, convex\, \,and}\, \, g(p)\le \mathcal{K}(p),p\in [0,1]\}, $$
$$\mathcal{K}^*=\inf\{g| g:[0,1]\rightarrow [0,1] \, {\rm  is\,\, concave\, \,and}\, \, g(p)\ge \mathcal{K}(p),p\in [0,1]\}. $$
Note that  $(-\mathcal{K})_*=-\mathcal{K}^*$. Moreover, if $\mathcal{K}$ is convex, then $\mathcal{K}_*=\mathcal{K}$; if $\mathcal{K}$ is concave, then $\mathcal{K}^*=\mathcal{K}$.
\end{definition}

Denote by $V(\mu,\sigma)$, $(\mu,\sigma)\in \mathbb{R}\times \mathbb{R}^+$, a set of random variables with mean $\mu$ and variance $\sigma^2$, namely
\[ V(\mu,\sigma)=\{X\in L^2(\Omega,\mathcal{F},\mathbb{P})\ |\ \mathbb{E}(X)=\mu\ \text{and}\ Var(X)=\sigma^2\}, \]
while $V_S(\mu,\sigma)$ denote a set of symmetric random variables with mean $\mu$ and variance $\sigma^2$, namely
\[ V_S(\mu,\sigma)=\{X\in L^2(\Omega,\mathcal{F},\mathbb{P})\ |\ \mathbb{E}(X)=\mu,  Var(X)=\sigma^2,\ \text{and}\ X\ \text{is\ symmetrical}\}. \]

For given distortion function $\mathcal{K}$,  we consider the
following    optimization problem (worst-case and best-case)
$$\sup_{X\in {\cal V}(\mu,\sigma)}\rho_{\mathcal{K}}[X]\,\, {\rm and} \,\, \inf_{X\in {\cal V}(\mu,\sigma)}\rho_{\mathcal{K}}[X],$$  respectively,
where ${\cal V}(\mu,\sigma)$ denotes either $V(\mu,\sigma)$ or $V_S(\mu,\sigma)$. If a random variable $X_*$  satisfies $\sup_{X\in {\cal V}(\mu,\sigma)}\rho_{\mathcal{K}}[X]=\rho_{\mathcal{K}}[X_*]$, then we refer to $F_{X_*}$ as a worst-case distribution. Similarly, we refer to $F_{X^*}$ as  a best-case distribution if  $\inf_{X\in {\cal V}(\mu,\sigma)}\rho_{\mathcal{K}}[X]=\rho_{\mathcal{K}}[X^*]$. The values  $\sup_{X\in {\cal V}(\mu,\sigma)}\rho_{\mathcal{K}}[X]$  and  $\inf_{X\in {\cal V}(\mu,\sigma)}\rho_{\mathcal{K}}[X]$   are correspondingly  the worst-case and best-case distortion risk measures, respectively.\\

\numberwithin{equation}{section}

\section{ Worst-case GlueVaR}
When the mean and variance of the underlying distributions are available, we first derive the worst-case GlueVaR distortion risk measure for all general distributions, and then consider, in addition, the symmetry shape factor to obtain the worst-case GlueVaR distortion risk measure.
\subsection {Case of general distributions}\label{intro}
Deriving special cases of worst-case distortion risk measures, such as worst-case VaR, worst-case TVaR, and worst-case RVaR, with known the first two moments of the underlying risks, is not new and results appear in various publications; see, e.g., Li (2018), Li et al. (2018), Zhu and Shao (2018), Cai et al. (2023), Shao and Zhang (2023a, 2023b) and Zhao et al. (2024). The following lemma is necessary for subsequent development.
\begin{lemma}[Shao and Zhang (2023a)] Let $X$ be a random variable with mean $\mu$ and
	variance $\sigma^2$   and $h$ be a
	distortion function. Then, the following statements hold:
	\begin{equation}
		\sup_{X\in V(\mu,\sigma)}\rho_{h}[X]=\mu+\sigma \sqrt{\int_0^1(\tilde{h}_*'(p)-1)^2dp},
	\end{equation}
	where $\tilde{h}_*'$	 denotes the right derivative function of $\tilde{h}_*$. Moreover, if $\tilde{h}_*'(p)= 1$ (a.e.),  the supremum in (3.1) is attained  by  any
	random variable $X\in V(\mu,\sigma)$;  if $\tilde{h}_*'(p)\neq 1$ (a.e.), the supremum in (3.1) is attained  by  the worst-case distribution of  rv $X_*$   with
	$$F_{X_*}^{-1+}(p)=\mu+\sigma \frac{\tilde{h}_*'(p)-1}{\sqrt{\int_0^1(\tilde{h}_*'(p)-1)^2dp}}.$$
\end{lemma}
In the sequel, we use Lemma 3.1 to derive the worst-case GlueVaR distortion risk measure and its attainable distribution.
\begin{proposition} Let $X$ be a random variable with mean $\mu$ and
	variance $\sigma^2$. Suppose GlueVaR has the distortion function $\mathcal{K}_{\beta,\alpha}^{h_1,h_2}$ defined by Eq. (2.1). Then, the following hold:\\
(i) When \( k_1\geq k_2 \) or \( k_1<k_2 \) and \(1-h_1\geq\frac{\beta-\alpha}{1-\alpha} \), we have
\begin{eqnarray}
\sup_{X\in V(\mu,\sigma)}\rho_{\mathcal{K}_{\beta,\alpha}^{h_1,h_2}}[X]=\mu+\sigma \sqrt{\frac{\alpha}{1-\alpha}, }
\end{eqnarray}
and the supremum in (3.2) is attained by the worst-case distribution of $X_*$ with	\begin{eqnarray}
		F_{X_*}^{-1+}(p)=\left\{\begin{array}{ll} \mu-\sigma \sqrt{\frac{1-\alpha}{\alpha}},  \ &{\rm if}\,\, p\in \left[0,\alpha\right],\\
			\mu+\sigma \sqrt{\frac{\alpha}{1-\alpha}}, \ &{\rm if}\, \,  p\in (\alpha,1].\\
		\end{array}
		\right.
\end{eqnarray}
(ii) When \( k_1<k_2 \) and \( 1-h_2\le1-h_1<\frac{\beta-\alpha}{1-\alpha} \), we have
\begin{eqnarray}
	\sup_{X\in V(\mu,\sigma)}\rho_{\mathcal{K}_{\beta,\alpha}^{h_1,h_2}}[X]=\mu+\sigma \sqrt{\alpha+\frac{(1-h_1-\beta+\alpha)^2}{\beta-\alpha}+\frac{(h_1-1+\beta)^2}{1-\beta}, }
\end{eqnarray}
and the supremum in (3.4) is attained by the worst-case distribution of $X_*$ with
\begin{eqnarray}
	F_{X_*}^{-1+}(p)=\left\{\begin{array}{ll}
		\mu-\sigma \sqrt{\frac{(\beta-\alpha)(1-\beta)}{(1-h_1)\eta+\beta(\beta-\alpha)}},  \ &{\rm if}\,\, p\in \left[0,\alpha\right],\\
		\mu+\sigma\frac{1-h_1-\beta+\alpha}{\sqrt{\beta-\alpha}}\sqrt{\frac{1-\beta}{(1-h_1)\eta+\beta(\beta-\alpha)}},  \ &{\rm if}\,\, p\in (\alpha,\beta],\\
		\mu+\sigma\frac{h_1-1+\beta}{\sqrt{1-\beta}}\sqrt{\frac{\beta-\alpha}{(1-h_1)\eta+\beta(\beta-\alpha)}}, \ &{\rm if}\, \,  p\in (\beta,1],\\
	\end{array}
	\right.
\end{eqnarray}
where $\eta=\alpha+\frac{(1-h_1-\beta+\alpha)^2}{\beta-\alpha}+\frac{(h_1-1+\beta)^2}{1-\beta}$.

\end{proposition}
{\bf Proof}.
From the distortion function $\mathcal{K}_{\beta,\alpha}^{h_1,h_2}$ in Equation (2.1), we directly get the following two expressions:
	\begin{eqnarray*}
	\mathcal{K}_{\beta,\alpha}^{h_1,h_2}(1-p)=\left\{\begin{array}{ll}
		1, \ &{\rm if}\, \,  p\in [0,\alpha],\\
		h_1+\frac{(h_2-h_1)(\beta-p)}{\beta-\alpha},  \ &{\rm if}\,\, p\in (\alpha,\beta],\\
		\frac{h_{1}(1-p)}{1-\beta}, \ &{\rm if}\, \,  p\in (\beta,1].\\
	\end{array}
	\right.
\end{eqnarray*}
and
\begin{eqnarray*}
	\widetilde{\mathcal{K}_{\beta,\alpha}^{h_1,h_2}}(p)=1-\mathcal{K}_{\beta,\alpha}^{h_1,h_2}(1-p)=\left\{\begin{array}{ll}
		0, \ &{\rm if}\, \,  p\in [0,\alpha],\\
		1-h_1-\frac{(h_2-h_1)(\beta-p)}{\beta-\alpha},  \ &{\rm if}\,\, p\in (\alpha,\beta],\\
		1-\frac{h_{1}(1-p)}{1-\beta}, \ &{\rm if}\, \,  p\in (\beta,1].\\
	\end{array}
	\right.
\end{eqnarray*}	

Since the size relationship between $k_1$ and $k_2$, i.e., the oblique relationship between the second and third segments of $\widetilde{\mathcal{K}_{\beta,\alpha}^{h_1,h_2}}(p)$, affects the convex envelope of $\widetilde{\mathcal{K}_{\beta,\alpha}^{h_1,h_2}}(p)$, we consider the following cases:\\
{\bf (i)} \( k_1\geq k_2 \) or \( k_1<k_2 \) and \(1-h_1\geq\frac{\beta-\alpha}{1-\alpha} \).

When \( k_1\geq k_2 \): Given the expressions for $k_1$ and $k_2$ above, we can easily get \( k_1\geq k_2\iff \frac{h_1}{h_2}\le \frac{\beta-1}{\alpha-1} \). In this case,
\begin{eqnarray*}
	\widetilde{\mathcal{K}_{\beta,\alpha}^{h_1,h_2}}_*(p)=\left\{\begin{array}{ll}
		0, \ &{\rm if}\, \,  p\in [0,\alpha],\\
		\frac{p-\alpha}{1-\alpha}, \ &{\rm if}\, \,  p\in (\alpha,1],\\
	\end{array}
	\right.
\end{eqnarray*}	
and
\begin{eqnarray*}
	\widetilde{\mathcal{K}_{\beta,\alpha}^{h_1,h_2}}_*^{'}(p)=\left\{\begin{array}{ll}
		0, \ &{\rm if}\, \,  p\in [0,\alpha],\\
		\frac{1}{1-\alpha}, \ &{\rm if}\, \,  p\in (\alpha,1].\\
	\end{array}
	\right.
\end{eqnarray*}	
Upon using Lemma 3.1, we get
\[ \int_0^1\left(\widetilde{\mathcal{K}_{\beta,\alpha}^{h_1,h_2}}_*^{'}(p)-1\right)^2dp=\frac{\alpha}{1-\alpha}. \]
So
\[ 	\sup_{X\in V(\mu,\sigma)}\rho_{\mathcal{K}_{\beta,\alpha}^{h_1,h_2}}[X]=\mu+\sigma \sqrt{\frac{\alpha}{1-\alpha},} \]
and the supremum in (3.2) is attained by the worst-case distribution of $X_*$ given by
(3.3).\\

 When \( k_1<k_2 \) and \(1-h_1\geq\frac{\beta-\alpha}{1-\alpha}\): We can easily get \( k_1<k_2\iff \frac{h_1}{h_2}> \frac{\beta-1}{\alpha-1} \). Because they have the same convex envelope, the proof for \( k_1<k_2 \) and \(1-h_1\geq\frac{\beta-\alpha}{1-\alpha}\) is similar to that for \( k_1\geq k_2\).\\
 
{\bf (ii)} $k_1<k_2$ and \(1-h_2\le1-h_1<\frac{\beta-\alpha}{1-\alpha}. \)\\
In this case,
\begin{eqnarray*}
	\widetilde{\mathcal{K}_{\beta,\alpha}^{h_1,h_2}}_*(p)=\left\{\begin{array}{ll}
		0, \ &{\rm if}\, \,  p\in [0,\alpha],\\
		\frac{(1-h_1)(p-\alpha)}{\beta-\alpha}, \ &{\rm if}\, \,  p\in (\alpha,\beta],\\
		\frac{h_1(p-1)}{1-\beta}+1, \ &{\rm if}\, \,  p\in (\beta,1],\\
	\end{array}
	\right.
\end{eqnarray*}	
and
\begin{eqnarray*}
	\widetilde{\mathcal{K}_{\beta,\alpha}^{h_1,h_2}}_*^{'}(p)=\left\{\begin{array}{ll}
		0, \ &{\rm if}\, \,  p\in [0,\alpha],\\
		\frac{1-h_1}{\beta-\alpha}, \ &{\rm if}\, \,  p\in (\alpha,\beta],\\
		\frac{h_1}{1-\beta}, \ &{\rm if}\, \,  p\in (\beta,1].\\
	\end{array}
	\right.
\end{eqnarray*}	
Upon using Lemma 3.1, we get
\[ \int_0^1\left(\widetilde{\mathcal{K}_{\beta,\alpha}^{h_1,h_2}}_*^{'}(p)-1\right)^2dp=\alpha+\frac{(1-h_1-\beta+\alpha)^2}{\beta-\alpha}+\frac{(h_1-1+\beta)^2}{1-\beta}. \]
So
\[ 	\sup_{X\in V(\mu,\sigma)}\rho_{\mathcal{K}_{\beta,\alpha}^{h_1,h_2}}[X]=\mu+\sigma \sqrt{\alpha+\frac{(1-h_1-\beta+\alpha)^2}{\beta-\alpha}+\frac{(h_1-1+\beta)^2}{1-\beta} }, \]
and the supremum in (3.4) is attained by the worst-case distribution of $X_*$ given by
(3.5). This ends the proof of Proposition 3.1.

 \begin{remark}
In Proposition 3.1, letting \(k_1=k_2=0\), \(k_1=0\) and \(k_2=\frac{1}{1-\alpha}\), \(k_1=\frac{1}{\beta-\alpha}\) and \(k_2=0\), respectively, we obtain the worst-case VaR, TVaR, and RVaR for general distributions as follows.
\[ 	\sup_{X\in V(\mu,\sigma)}\rho_{\mathcal{K}_{\alpha,\alpha}^{0,0}}[X]=\sup_{X\in V(\mu,\sigma)}\rho_{\mathcal{K}_{\alpha,\alpha}^{1,1}}[X]=\sup_{X\in V(\mu,\sigma)}\rho_{\mathcal{K}_{\beta,\alpha}^{0,1}}[X]=\mu+\sigma \sqrt{\frac{\alpha}{1-\alpha} }, \]
and the supremum is attained by the worst-case distribution of $X_*$ with
\begin{eqnarray*}
	F_{X_*}^{-1+}(p)=\left\{\begin{array}{ll} \mu-\sigma \sqrt{\frac{1-\alpha}{\alpha}},  \ &{\rm if}\,\, p\in \left[0,\alpha\right],\\
		\mu+\sigma \sqrt{\frac{\alpha}{1-\alpha}}, \ &{\rm if}\, \,  p\in (\alpha,1].\\
	\end{array}
	\right.
\end{eqnarray*}

Obviously, the above results are consistent with findings in the literature, see, e.g., Li (2018), Li et al. (2018), Zhu and Shao (2018), Bernard et al. (2020), Cai et al. (2023), Shao and
Zhang (2023a, 2023b) and Zhao et al. (2024).
 \end{remark}
 \begin{remark}
The worst-case of GlueVaR distortion risk measure in conclusion (ii) of proposition 3.1 and its corresponding attainable distribution are consistent with the results of Shao and
Zhang (2023b).
 \end{remark}
\subsection {Case of symmetric distributions}\label{intro}
With the first two moments and symmetry of the underlying distributions are known, Li et al. (2018) and Zhu and Shao (2018) studied the worst-case RVaR and worst-case DRM and their corresponding distributions, respectively. Next, we derive the worst-case GlueVaR distortion risk measure and its corresponding distribution when the first two moments and symmetry of the underlying distributions are known.
\begin{lemma}[Shao and Zhang (2023a)] Let $X$ be a  symmetric random variable with mean $\mu$ and
	variance $\sigma^2$ and  $h$ be a
	distortion function. Then, the following statements hold:
	\begin{equation}
		\sup_{X\in V_S(\mu,\sigma)}\rho_{h}[X]=\mu+\frac12\sigma \sqrt{\int_0^1(\tilde{h}_*'(p)-\tilde{h}_*'(1-p))^2dp},
	\end{equation}
	where $\tilde{h}_*'$	 denotes the right derivative function of $\tilde{h}_*$. Moreover, if  $\tilde{h}_*'(p)-\tilde{h}_*'(1-p)=0$ (a.e.),  the supremum in (3.6) is attained  by  any random variable $X\in V_S(\mu,\sigma)$;  if  $\tilde{h}_*'(p)-\tilde{h}_*'(1-p)\neq 0$ (a.e.), the supremum in (3.6) is attained  by the  the worst-case distribution of  rv $X_*$   with
	$$F_{X_*}^{-1+}(p)=\mu+\sigma \frac{\tilde{h}_*'(p)-\tilde{h}_*'(1-p)}{\sqrt{\int_0^1(\tilde{h}_*'(p)-\tilde{h}_*'(1-p))^2dp}}.$$
\end{lemma}
Based on the above lemma, we have the following proposition.
\begin{proposition}
	Let $X$ be a symmetric random variable with mean $\mu$ and
	variance $\sigma^2$. Suppose GlueVaR has the distortion function $\mathcal{K}_{\beta,\alpha}^{h_1,h_2}$ defined by Eq. (2.1). Then, the following statements hold for $\frac{1}{2}\le\alpha<\beta< 1 $:\\
(i) When \( k_1\geq k_2 \) or \( k_1<k_2 \) and \( 1-h_1\geq\frac{\beta-\alpha}{1-\alpha} \), we have
	\begin{eqnarray}
		\sup_{X\in V_S(\mu,\sigma)}\rho_{\mathcal{K}_{\beta,\alpha}^{h_1,h_2}}[X]=\mu+\sigma \sqrt{\frac{1}{2(1-\alpha)}, }
	\end{eqnarray}
	if  $\widetilde{\mathcal{K}_{\beta,\alpha}^{h_1,h_2}}_*^{'}(p)-\widetilde{\mathcal{K}_{\beta,\alpha}^{h_1,h_2}}_*^{'}(1-p)=0$ (a.e.), i.e., $p\in (1-\alpha,\alpha)$, the supremum in (3.7) is attained  by  any random variable $X\in V_S(\mu,\sigma)$; if  $\widetilde{\mathcal{K}_{\beta,\alpha}^{h_1,h_2}}_*^{'}(p)-\widetilde{\mathcal{K}_{\beta,\alpha}^{h_1,h_2}}_*^{'}(1-p)\neq 0$ (a.e.), the supremum in (3.7) is attained by  the worst-case distribution of  rv $X_*$  with
	\begin{eqnarray}
		F_{X_*}^{-1+}(p)=\left\{\begin{array}{ll} \mu-\sigma \sqrt{\frac{1}{2(1-\alpha)} },  \ &{\rm if}\,\, p\in \left[0,1-\alpha\right),\\
			\mu+\sigma \sqrt{\frac{1}{2(1-\alpha)} }, \ &{\rm if}\, \,  p\in [\alpha,1].\\
		\end{array}
		\right.
	\end{eqnarray}
(ii) When \( k_1<k_2 \) and \( 1-h_2\le1-h_1<\frac{\beta-\alpha}{1-\alpha} \), we have
\begin{eqnarray}
	\sup_{X\in V_S(\mu,\sigma)}\rho_{\mathcal{K}_{\beta,\alpha}^{h_1,h_2}}[X]=\mu+\sigma \sqrt{\frac{q}{2(1-\beta)(\beta-\alpha)},}
\end{eqnarray}
if  $\widetilde{\mathcal{K}_{\beta,\alpha}^{h_1,h_2}}_*^{'}(p)-\widetilde{\mathcal{K}_{\beta,\alpha}^{h_1,h_2}}_*^{'}(1-p)=0$ (a.e.), i.e., $p\in (1-\alpha,\alpha)$, the supremum in (3.9) is attained  by  any
random variable $X\in V_S(\mu,\sigma)$; if  $\widetilde{\mathcal{K}_{\beta,\alpha}^{h_1,h_2}}_*^{'}(p)-\widetilde{\mathcal{K}_{\beta,\alpha}^{h_1,h_2}}_*^{'}(1-p)\neq 0$ (a.e.), the supremum in (3.9) is attained  by   the worst-case distribution of  rv $X_*$  with
\begin{eqnarray}
	F_{X_*}^{-1+}(p)=\left\{\begin{array}{ll}
		\mu-\sigma h_1 \sqrt{\frac{\beta-\alpha}{2(1-\beta)\zeta} }, \ &{\rm if}\,\, p\in \left[0,1-\beta\right),\\
		\mu-\sigma (1-h_1) \sqrt{\frac{1-\beta}{2(\beta-\alpha)\zeta} }, \ &{\rm if}\,\, p\in \left[1-\beta,1-\alpha\right),\\
		\mu+\sigma (1-h_1) \sqrt{\frac{1-\beta}{2(\beta-\alpha)\zeta} }, \ &{\rm if}\,\, p\in \left[\alpha,\beta\right),\\
		\mu+\sigma h_1 \sqrt{\frac{\beta-\alpha}{2(1-\beta)\zeta} }, \ &{\rm if}\,\, p\in \left[\beta,1\right],\\
	\end{array}
	\right.
\end{eqnarray}
where $\zeta=h_1^2(1-\alpha)+(1-2h_1)(1-\beta)$.
\end{proposition}
{\bf Proof}. Using the same arguments as Proposition 3.1, we consider the following classification of cases consistent with Proposition 3.1:\\
{\bf (i)} \( k_1\geq k_2 \) or \( k_1<k_2 \) and \( 1-h_1\geq\frac{\beta-\alpha}{1-\alpha} \).\\
In this case,
	\begin{eqnarray*}
		\widetilde{\mathcal{K}_{\beta,\alpha}^{h_1,h_2}}_*^{'}(p)=\left\{\begin{array}{ll}
			0, \ &{\rm if}\, \,  p\in [0,\alpha],\\
			\frac{1}{1-\alpha}, \ &{\rm if}\, \,  p\in (\alpha,1].\\
		\end{array}
		\right.
	\end{eqnarray*}	
	and
	\begin{eqnarray*}
		\widetilde{\mathcal{K}_{\beta,\alpha}^{h_1,h_2}}_*^{'}(1-p)=\left\{\begin{array}{ll}
			\frac{1}{1-\alpha}, \ &{\rm if}\, \,  p\in [0,1-\alpha),\\
			0, \ &{\rm if}\, \,  p\in [1-\alpha,1].\\
		\end{array}
		\right.
	\end{eqnarray*}	
	Upon using Lemma 3.2, we get
	\[ \int_0^1\left(\widetilde{\mathcal{K}_{\beta,\alpha}^{h_1,h_2}}_*^{'}(p)-\widetilde{\mathcal{K}_{\beta,\alpha}^{h_1,h_2}}_*^{'}(1-p)\right)^2dp=\frac{2}{1-\alpha}. \]
	So
	\[ 	\sup_{X\in V_S(\mu,\sigma)}\rho_{\mathcal{K}_{\beta,\alpha}^{h_1,h_2}}[X]=\mu+\sigma \sqrt{\frac{1}{2(1-\alpha)} }, \]
	if  $\widetilde{\mathcal{K}_{\beta,\alpha}^{h_1,h_2}}_*^{'}(p)-\widetilde{\mathcal{K}_{\beta,\alpha}^{h_1,h_2}}_*^{'}(1-p)=0$ (a.e.), i.e., $p\in (1-\alpha,\alpha)$, the supremum in (3.7) is attained  by  any random variable $X\in V_S(\mu,\sigma)$; if  $\widetilde{\mathcal{K}_{\beta,\alpha}^{h_1,h_2}}_*^{'}(p)-\widetilde{\mathcal{K}_{\beta,\alpha}^{h_1,h_2}}_*^{'}(1-p)\neq 0$ (a.e.), the supremum in (3.7) is attained  by the worst-case distribution of  rv $X_*$  with (3.8).\\
{\bf (ii)} $k_1<k_2$ and \(1-h_2\le1-h_1<\frac{\beta-\alpha}{1-\alpha} \).\\
In this case,	
	\begin{eqnarray*}
		\widetilde{\mathcal{K}_{\beta,\alpha}^{h_1,h_2}}_*^{'}(p)=\left\{\begin{array}{ll}
			0, \ &{\rm if}\, \,  p\in [0,\alpha],\\
			\frac{1-h_1}{\beta-\alpha}, \ &{\rm if}\, \,  p\in (\alpha,\beta],\\
			\frac{h_1}{1-\beta}, \ &{\rm if}\, \,  p\in (\beta,1],\\
		\end{array}
		\right.
	\end{eqnarray*}
	and
	\begin{eqnarray*}
		\widetilde{\mathcal{K}_{\beta,\alpha}^{h_1,h_2}}_*^{'}(1-p)=\left\{\begin{array}{ll}
			\frac{h_1}{1-\beta}, \ &{\rm if}\, \,  p\in[0,1-\beta) ,\\
			\frac{1-h_1}{\beta-\alpha}, \ &{\rm if}\, \,  p\in [1-\beta,1-\alpha),\\
			0, \ &{\rm if}\, \,  p\in [1-\alpha,1].\\
		\end{array}
		\right.
	\end{eqnarray*}	
Applying Lemma 3.2, we get
\begin{eqnarray*}
\int_0^1\left(\widetilde{\mathcal{K}_{\beta,\alpha}^{h_1,h_2}}_*^{'}(p)-\widetilde{\mathcal{K}_{\beta,\alpha}^{h_1,h_2}}_*^{'}(1-p)\right)^2dp=\frac{2h_1^2}{1-\beta}+\frac{2(1-h_1)^2}{\beta-\alpha}=\frac{2\zeta}{(1-\beta)(\beta-\alpha)},
\end{eqnarray*}
where $\zeta=h_1^2(1-\alpha)+(1-2h_1)(1-\beta)$.\\
So
	\[ 	\sup_{X\in V_S(\mu,\sigma)}\rho_{\mathcal{K}_{\beta,\alpha}^{h_1,h_2}}[X]=\mu+\sigma \sqrt{\frac{\zeta}{2(1-\beta)(\beta-\alpha)} }, \]
	if  $\widetilde{\mathcal{K}_{\beta,\alpha}^{h_1,h_2}}_*^{'}(p)-\widetilde{\mathcal{K}_{\beta,\alpha}^{h_1,h_2}}_*^{'}(1-p)=0$ (a.e.), i.e., $p\in (1-\alpha,\alpha)$, the supremum in (3.9) is attained by any random variable $X\in V_S(\mu,\sigma)$; if  $\widetilde{\mathcal{K}_{\beta,\alpha}^{h_1,h_2}}_*^{'}(p)-\widetilde{\mathcal{K}_{\beta,\alpha}^{h_1,h_2}}_*^{'}(1-p)\neq 0$ (a.e.), the supremum in (3.9) is attained by the worst-case distribution of rv $X_*$ with (3.10). This ends the proof of Proposition 3.2.
 \begin{remark}
For $0<\alpha<\beta< \frac{1}{2}$, we can express $\rho_{\mathcal{K}_{\beta,\alpha}^{h_1,h_2}}[X]$ as
	\begin{eqnarray*}
	\rho_{\mathcal{K}_{\beta,\alpha}^{h_1,h_2}}[X]&&=\int_0^1 F_X^{-1}(1-p)d{\mathcal{K}_{\beta,\alpha}^{h_1,h_2}}(p)\\&&\le\mu\int_0^1d\mathcal{K}_{\beta,\alpha}^{h_1,h_2}(p)\\&&=\mu.
	\end{eqnarray*}
	We thus conclude that $\sup_{X\in V_S(\mu,\sigma)}\rho_{\mathcal{K}_{\beta,\alpha}^{h_1,h_2}}[X]=\mu $, for $0<\alpha<\beta< \frac{1}{2} $.
\end{remark}
\begin{remark}Using Proposition 3.2, we get the worst-case VaR, TVaR, and RVaR for symmetric distributions as follows.\\
(i) For $0<\alpha<\beta<\frac{1}{2}$, we have
\[\sup_{X\in V_S(\mu,\sigma)}\rho_{\mathcal{K}_{\alpha,\alpha}^{0,0}}[X]=\sup_{X\in V_S(\mu,\sigma)}\rho_{\mathcal{K}_{\beta,\alpha}^{0,1}}[X]=\mu, \]
and the supremum is attained by any random variable $X_*\in V_S(\mu,\sigma)$.\\
(ii) For $\frac{1}{2}\le\alpha<\beta<1$, we have
  \begin{eqnarray}
\sup_{X\in V_S(\mu,\sigma)}\rho_{\mathcal{K}_{\alpha,\alpha}^{0,0}}[X]&&=\sup_{X\in V_S(\mu,\sigma)}\rho_{\mathcal{K}_{\alpha,\alpha}^{1,1}}[X]\nonumber\\&&=\sup_{X\in V_S(\mu,\sigma)}\rho_{\mathcal{K}_{\beta,\alpha}^{0,1}}[X]=\mu+\sigma \sqrt{\frac{1}{2(1-\alpha)}, }
 \end{eqnarray}
 if  $\widetilde{\mathcal{K}_{\beta,\alpha}^{h_1,h_2}}_*^{'}(p)-\widetilde{\mathcal{K}_{\beta,\alpha}^{h_1,h_2}}_*^{'}(1-p)=0$ (a.e.), i.e., $p\in (1-\alpha,\alpha)$, the supremum in (3.11) is attained  by  any
 random variable $X\in V_S(\mu,\sigma)$; if  $\widetilde{\mathcal{K}_{\beta,\alpha}^{h_1,h_2}}_*^{'}(p)-\widetilde{\mathcal{K}_{\beta,\alpha}^{h_1,h_2}}_*^{'}(1-p)\neq 0$ (a.e.), the supremum in (3.11) is attained  by   the worst-case distribution of  rv $X_*$  with
 \begin{eqnarray*}
 	F_{X_*}^{-1+}(p)=\left\{\begin{array}{ll} \mu-\sigma \sqrt{\frac{1}{2(1-\alpha)} },  \ &{\rm if}\,\, p\in \left[0,1-\alpha\right),\\
 		\mu+\sigma \sqrt{\frac{1}{2(1-\alpha)} }, \ &{\rm if}\, \,  p\in [\alpha,1],\\
 	\end{array}
 	\right.
 \end{eqnarray*}
where $\mathcal{K}_{\beta,\alpha}^{h_1,h_2}(p)$ is, respectively, $\mathcal{K}_{\alpha,\alpha}^{0,0}(p)$, $\mathcal{K}_{\alpha,\alpha}^{1,1}(p)$ and $\mathcal{K}_{\beta,\alpha}^{0,1}(p)$.

 The results in the above coincide with those in Shao and
 Zhang (2023a, 2023b), Zhao et al. (2024).
\end{remark}

\section{ Best-case GlueVaR}
In this section, we consider the best-case GlueVaR and compute their explicit expressions with knowledge of the first two moments and symmetry of the underlying distributions.
\subsection {Case of general distributions}\label{intro}
There has been much literature that derives the best-case of special distortion risk measures when the first two moments of the underlying distributions are known, such as best-case VaR, best-case TVaR, best-case RVaR and so on. See, for example, Zhu and Shao (2018), Shao and Zhang (2023a, 2023b) and Zhao et al. (2024). We use the following lemma to obtain the best-case GlueVaR distortion risk measure for general random variables.
\begin{lemma}[Shao and Zhang (2023a)]
Let $X$ be a random variable with mean $\mu$ and
variance $\sigma^2$   and $h$ be a
distortion function. Then, the following statements hold:
\begin{eqnarray*}
	\inf_{X\in V(\mu,\sigma)}\rho_{h}[X]= \mu-\sigma  \sqrt{\int_0^1({h}_*'(p)-1)^2dp}.
\end{eqnarray*}
The best-case rv $X^*= \arg\inf_{X\in V(\mu,\sigma)}\rho_{h}[X] $ is  with
\begin{eqnarray*}
	F_{X^*}^{-1+}(p)=\left\{\begin{array}{ll} \mu,  \ &{\rm if}\,\, h_*'(p)= 1 \ (a.e.),\\
		\mu-\sigma \frac{{h}_*'(p)-1}{\sqrt{\int_0^1({h}_*'(p)-1)^2dp}}, \ &{\rm if}\, \,  h_*'(p)\neq 1 \ (a.e.).\\
	\end{array}
	\right.
\end{eqnarray*}
\end{lemma}
We have the following proposition.
\begin{proposition} Let $X$ be a random variable with mean $\mu$ and
	variance $\sigma^2$. Suppose GlueVaR has the distortion function $\mathcal{K}_{\beta,\alpha}^{h_1,h_2}$ defined by Eq. (2.1). Then, the following hold:\\
	(i) When \( k_2\geq k_1 \) and \( h_2\geq 1-\alpha \) or $\max\{k_2,k_3\}<k_1$ and $k_2\geq1$, we have
	\begin{eqnarray*}
		\inf_{X\in V(\mu,\sigma)}\rho_{\mathcal{K}_{\beta,\alpha}^{h_1,h_2}}[X]=\mu,
	\end{eqnarray*}	
	and the best-case rv $X^*= \arg\inf_{X\in V(\mu,\sigma)}\rho_{\mathcal{K}_{\beta,\alpha}^{h_1,h_2}}[X]$ is  with
	\begin{eqnarray}
		F_{X^*}^{-1+}(p)=\mu, p\in[0,1].
	\end{eqnarray}	
	(ii) When \( k_2\geq k_1 \) and \( h_2<1-\alpha \), we have
	\begin{eqnarray*}
		\inf_{X\in V(\mu,\sigma)}\rho_{\mathcal{K}_{\beta,\alpha}^{h_1,h_2}}[X]=\mu+\sigma\frac{h_2-1+\alpha}{\sqrt{\alpha(1-\alpha)}},
	\end{eqnarray*}	
	and the best-case rv $X^*= \arg\inf_{X\in V(\mu,\sigma)}\rho_{\mathcal{K}_{\beta,\alpha}^{h_1,h_2}}[X]$ is  with
	\begin{eqnarray}
		F_{X_*}^{-1+}(p)=\left\{\begin{array}{ll}
			\mu+\sigma \sqrt{\frac{\alpha}{1-\alpha}}, \ &{\rm if}\, \,  p\in \left[0,1-\alpha\right),\\
			\mu-\sigma \sqrt{\frac{1-\alpha}{\alpha}},  \ &{\rm if}\,\, p\in [1-\alpha,1].\\
		\end{array}
		\right.
	\end{eqnarray}
	(iii) When \( k_3>k_1>k_2 \), we have
	\begin{eqnarray*}
		\inf_{X\in V(\mu,\sigma)}\rho_{\mathcal{K}_{\beta,\alpha}^{h_1,h_2}}[X]=\mu-\sigma\sqrt{\xi},
	\end{eqnarray*}
	and the best-case rv $X^*= \arg\inf_{X\in V(\mu,\sigma)}\rho_{\mathcal{K}_{\beta,\alpha}^{h_1,h_2}}[X]$ is  with
	\begin{eqnarray}
		F_{X^*}^{-1+}(p)=\left\{\begin{array}{ll}
			\mu-\sigma\frac{h_1-1+\beta}{(1-\beta)\sqrt{\xi}}, \ &{\rm if}\, \,  p\in [0,1-\beta),\\
			\mu-\sigma\frac{h_2-h_1-\beta+\alpha}{(\beta-\alpha)\sqrt{\xi}}, \ &{\rm if}\, \,  p\in [1-\beta,1-\alpha),\\
			\mu-\sigma\frac{1-h_2-\alpha}{\alpha\sqrt{\xi}}, \ &{\rm if}\, \,  p\in [1-\alpha,1],\\
		\end{array}
		\right.
	\end{eqnarray}
	where $\xi=\frac{(1-h_2-\alpha)^2}{\alpha}+\frac{(h_2-h_1-\beta+\alpha)^2}{\beta-\alpha}+\frac{(h_1-1+\beta)^2}{1-\beta}$.\\
	(iv) When \( k_3=k_1>k_2 \) or $\max\{k_2,k_3\}<k_1$ and $k_2<1$, we have
	\begin{eqnarray*}
		\inf_{X\in V(\mu,\sigma)}\rho_{\mathcal{K}_{\beta,\alpha}^{h_1,h_2}}[X]=\mu+\sigma\frac{h_1-1+\beta}{\sqrt{\beta(1-\beta)}},
	\end{eqnarray*}
	and the best-case rv $X^*= \arg\inf_{X\in V(\mu,\sigma)}\rho_{\mathcal{K}_{\beta,\alpha}^{h_1,h_2}}[X]$ is  with
	\begin{eqnarray}
		F_{X^*}^{-1+}(p)=\left\{\begin{array}{ll}
			\mu+\sigma \sqrt{\frac{\beta}{1-\beta}}, \ &{\rm if}\, \,  p\in \left[0,1-\beta\right),\\
			\mu-\sigma \sqrt{\frac{1-\beta}{\beta}},  \ &{\rm if}\,\, p\in [1-\beta,1].\\
			
		\end{array}
		\right.
	\end{eqnarray}
\end{proposition}
{\bf Proof}.
According to the magnitude relationship between the slopes, we will divide it into the following cases:\\
{\bf (i)} \( k_2\geq k_1 \) and \( h_2\geq 1-\alpha \) or $\max\{k_2,k_3\}<k_1$ and $k_2\geq1$.

When $k_2\geq k_1$ and $h_2\geq 1-\alpha$:
Substituting the expressions for $k_1$ and $k_2$, we get \( k_2\geq k_1\iff \frac{h_1}{h_2}\geq \frac{\beta-1}{\alpha-1} \). With these conditions, we have
\[ 	{\mathcal{K}_{\beta,\alpha}^{h_1,h_2}}_*(p)=p,p\in[0,1] \]
and
\[ 	{\mathcal{K}_{\beta,\alpha}^{h_1,h_2}}_*^{'}(p)=1, p\in[0,1]. \]
By using Lemma 4.1, we get
\[ \int_0^1\left({\mathcal{K}_{\beta,\alpha}^{h_1,h_2}}_*^{'}(p)-1\right)^2dp=0. \]
So
\[ 	\inf_{X\in V(\mu,\sigma)}\rho_{\mathcal{K}_{\beta,\alpha}^{h_1,h_2}}[X]=\mu,  \]
and the best-case rv $X^*= \arg\inf_{X\in V(\mu,\sigma)}\rho_{\mathcal{K}_{\beta,\alpha}^{h_1,h_2}}[X]$ is  with (4.1).

When $\max\{k_2,k_3\}<k_1$ and $k_2\geq1$: Because they have the same convex envelope, the proof is similar to that for \( k_2\geq k_1\) and \( h_2\geq 1-\alpha \).\\
 
{\bf (ii)} $k_2\geq k_1$ and $h_2< 1-\alpha$.\\
In this case,
\begin{eqnarray*}
	{\mathcal{K}_{\beta,\alpha}^{h_1,h_2}}_*(p)=\left\{\begin{array}{ll}
		\frac{h_2p}{1-\alpha}, \ &{\rm if}\, \,  p\in [0,1-\alpha),\\
		\frac{(1-h_2)(p-1)}{\alpha}+1, \ &{\rm if}\, \,  p\in [1-\alpha,1].\\
	\end{array}
	\right.
\end{eqnarray*}	
and
\begin{eqnarray*}
	{\mathcal{K}_{\beta,\alpha}^{h_1,h_2}}_*^{'}(p)=\left\{\begin{array}{ll}
		\frac{h_2}{1-\alpha}, \ &{\rm if}\, \,  p\in [0,1-\alpha),\\
		\frac{1-h_2}{\alpha}, \ &{\rm if}\, \,  p\in [1-\alpha,1].\\
	\end{array}
	\right.
\end{eqnarray*}	
Applying Lemma 4.1, we get
\[ \int_0^1\left({\mathcal{K}_{\beta,\alpha}^{h_1,h_2}}_*^{'}(p)-1\right)^2dp=\frac{(h_2-1+\alpha)^2}{\alpha(1-\alpha)}. \]
So
\[ 	\inf_{X\in V(\mu,\sigma)}\rho_{\mathcal{K}_{\beta,\alpha}^{h_1,h_2}}[X]=\mu-\sigma\sqrt{\frac{(h_2-1+\alpha)^2}{\alpha(1-\alpha)} }=\mu+\sigma\frac{h_2-1+\alpha}{\sqrt{\alpha(1-\alpha)}},\]
and the best-case rv $X^*= \arg\inf_{X\in V(\mu,\sigma)}\rho_{\mathcal{K}_{\beta,\alpha}^{h_1,h_2}}[X]$ is  with (4.2).

{\bf (iii)} $k_2< k_1<k_3$.

Based on the expressions of $k_1$ and $k_2$, we get \(k_2<k_1\iff \frac{h_1}{h_2}< \frac{\beta-1}{\alpha-1} \). When $k_2< k_1<k_3$, we have
\begin{eqnarray*}
	{\mathcal{K}_{\beta,\alpha}^{h_1,h_2}}_*(p)=\left\{\begin{array}{ll}
		\frac{h_1p}{1-\beta}, \ &{\rm if}\, \,  p\in [0,1-\beta),\\
		h_1+\frac{(h_2-h_1)[p-(1-\beta)]}{\beta-\alpha}, \ &{\rm if}\, \,  p\in [1-\beta,1-\alpha),\\
		\frac{(1-h_2)(p-1)}{\alpha}+1, \ &{\rm if}\, \,  p\in [1-\alpha,1].\\
	\end{array}
	\right.
\end{eqnarray*}	
and
\begin{eqnarray*}
	{\mathcal{K}_{\beta,\alpha}^{h_1,h_2}}_*^{'}(p)=\left\{\begin{array}{ll}
		\frac{h_1}{1-\beta}, \ &{\rm if}\, \,  p\in [0,1-\beta),\\
		h_1+\frac{h_2-h_1}{\beta-\alpha}, \ &{\rm if}\, \,  p\in [1-\beta,1-\alpha),\\
		\frac{1-h_2}{\alpha}+1, \ &{\rm if}\, \,  p\in [1-\alpha,1].\\
	\end{array}
	\right.
\end{eqnarray*}	
By using Lemma 4.1, we get
\begin{eqnarray*}
	\int_0^1\left({\mathcal{K}_{\beta,\alpha}^{h_1,h_2}}_*^{'}(p)-1\right)^2dp=\xi,
\end{eqnarray*}	
where $\xi=\frac{(1-h_2-\alpha)^2}{\alpha}+\frac{(h_2-h_1-\beta+\alpha)^2}{\beta-\alpha}+\frac{(h_1-1+\beta)^2}{1-\beta}$.\\
So
\begin{eqnarray*}
	\inf_{X\in V(\mu,\sigma)}\rho_{\mathcal{K}_{\beta,\alpha}^{h_1,h_2}}[X]=\mu-\sigma\sqrt{\xi},
\end{eqnarray*}
and the best-case rv $X^*= \arg\inf_{X\in V(\mu,\sigma)}\rho_{\mathcal{K}_{\beta,\alpha}^{h_1,h_2}}[X]$ is  with (4.3).

{\bf (iv)} \(k_3=k_1>k_2\) or $\max\{k_2,k_3\}<k_1$ and $k_2<1$.

When \(k_3=k_1>k_2\): It is easy to see that not only $k_3=k_1>k_2$ but also $\frac{h_1}{1-\beta}<1 $, and $\frac{h_1}{1-\beta}<1 $ plays a role in simplifying the expression further. Under these conditions, we have
\begin{eqnarray*}
	{\mathcal{K}_{\beta,\alpha}^{h_1,h_2}}_*(p)=\left\{\begin{array}{ll}
		\frac{h_1p}{1-\beta}, \ &{\rm if}\, \,  p\in [0,1-\beta),\\
		\frac{(1-h_1)(p-1)}{\beta}+1, \ &{\rm if}\, \,  p\in [1-\beta,1].\\
	\end{array}
	\right.
\end{eqnarray*}	
and
\begin{eqnarray*}
	{\mathcal{K}_{\beta,\alpha}^{h_1,h_2}}_*^{'}(p)=\left\{\begin{array}{ll}
		\frac{h_1}{1-\beta}, \ &{\rm if}\, \,  p\in [0,1-\beta),\\
		\frac{1-h_1}{\beta}, \ &{\rm if}\, \,  p\in [1-\beta,1].\\
	\end{array}
	\right.
\end{eqnarray*}	
By using Lemma 4.1, we get
\[ \int_0^1\left({\mathcal{K}_{\beta,\alpha}^{h_1,h_2}}_*^{'}(p)-1\right)^2dp=\frac{(h_1-1+\beta)^2}{\beta(1-\beta)}. \]
So
\[ 	\inf_{X\in V(\mu,\sigma)}\rho_{\mathcal{K}_{\beta,\alpha}^{h_1,h_2}}[X]=\mu-\sigma\sqrt{\frac{(h_1-1+\beta)^2}{\beta(1-\beta)} }=\mu+\sigma\frac{h_1-1+\beta}{\sqrt{\beta(1-\beta)}},\]
and the best-case rv $X^*= \arg\inf_{X\in V(\mu,\sigma)}\rho_{\mathcal{K}_{\beta,\alpha}^{h_1,h_2}}[X]$ is  with (4.4).

When $\max\{k_2,k_3\}<k_1$ and $k_2<1$: The proof is similar to that for $k_2<k_1=k_3$. This completes the proof of Proposition 4.1.\\
 
\begin{remark}
Given $k_1$ and $k_2$ in Remark 3.1, let $k_3$ be \(\frac{1}{\alpha}\), \(0\), and \(0\), respectively, in Proposition 4.1, we obtain the best-case VaR, TVaR, and RVaR for general distributions. For RVaR, we have
\begin{eqnarray}
\inf_{X\in V(\mu,\sigma)}RVaR_{\alpha,\beta}[X]=\inf_{X\in V(\mu,\sigma)}\rho_{\mathcal{K}_{\beta,\alpha}^{0,1}}[X]=\mu-\sigma\sqrt{\frac{1-\beta}{\beta}},
\end{eqnarray}
and the best-case rv $X^*= \arg\inf_{X\in V(\mu,\sigma)}\rho_{\mathcal{K}_{\beta,\alpha}^{0,1}}[X]$ is  with
\begin{eqnarray}
	F_{X^*}^{-1+}(p)=\left\{\begin{array}{ll}
		\mu+\sigma \sqrt{\frac{\beta}{1-\beta}}, \ &{\rm if}\, \,  p\in \left[0,1-\beta\right),\\
		\mu-\sigma \sqrt{\frac{1-\beta}{\beta}},  \ &{\rm if}\,\, p\in [1-\beta,1].\\
		
	\end{array}
	\right.
\end{eqnarray}

In particular, letting $\beta\to \alpha$ in (4.5) and (4.6), respectively, we obtain
\[ 	\inf_{X\in V(\mu,\sigma)}VaR_{\alpha}[X]=\inf_{X\in V(\mu,\sigma)}\rho_{\mathcal{K}_{\beta,\alpha}^{0,0}}[X]=\mu-\sigma\sqrt{\frac{1-\alpha}{\alpha} },\]
and the best-case rv $X^*= \arg\inf_{X\in V(\mu,\sigma)}\rho_{\mathcal{K}_{\beta,\alpha}^{0,0}}[X]$ is  with
\begin{eqnarray*}
	F_{X^*}^{-1+}(p)=\left\{\begin{array}{ll}
		\mu+\sigma \sqrt{\frac{\alpha}{1-\alpha}}, \ &{\rm if}\, \,  p\in \left[0,1-\alpha\right),\\
		\mu-\sigma \sqrt{\frac{1-\alpha}{\alpha}},  \ &{\rm if}\,\, p\in [1-\alpha,1].\\
		
	\end{array}
	\right.
\end{eqnarray*}

Additionally, letting $\beta\to 1$ in (4.5) and (4.6), respectively, we get
\[ 	\inf_{X\in V(\mu,\sigma)}TVaR_{\alpha}[X]=\inf_{X\in V(\mu,\sigma)}\rho_{\mathcal{K}_{\alpha,\alpha}^{1,1}}[X]=\mu,  \]
and the best-case rv $X^*= \arg\inf_{X\in V(\mu,\sigma)}\rho_{\mathcal{K}_{\alpha,\alpha}^{1,1}}[X]$ is  with
\[ 	F_{X^*}^{-1+}(p)=\mu, p\in[0,1]. \]

The results are obviously in line with that of Zhao et al. (2024) and Zhu and Shao (2018), among others.
\end{remark}
\subsection {Case of symmetric distributions}\label{intro}
For symmetric distributions, given the first two moments, we apply the following lemma to derive the best-case GlueVaR distortion risk measure and its associated distribution.
\begin{lemma}[Shao and Zhang (2023a)]	Let $X$ be a  symmetric random variable with mean $\mu$ and
	variance $\sigma^2$ and  $h$ be a
	distortion function. Then, the following statements hold:
	\begin{eqnarray*}
		\inf_{X\in V_S(\mu,\sigma)}\rho_{h}[X]=  \mu-\frac12\sigma \sqrt{\int_0^1({h}_*'(p)-{h}_*'(1-p))^2dp}.
	\end{eqnarray*}
	The best-case rv $X^*= \arg\inf_{X\in V_S(\mu,\sigma)}\rho_{h}[X] $  is  with
	\begin{eqnarray*}
		F_{X^*}^{-1+}(p)=\left\{\begin{array}{ll} \mu,  \ &{\rm if}\,\,  {h}_*'(p)-{h}_*'(1-p)=0 \ (a.e.),\\
			\mu-\sigma \frac{{h}_*'(p)-{h}_*'(1-p)}{\sqrt{\int_0^1({h}_*'(p)-{h}_*'(1-p))^2dp}}, \ &{\rm if}\, \,  {h}_*'(p)-{h}_*'(1-p) \neq 0  \ (a.e.).\\
		\end{array}
		\right.
	\end{eqnarray*}
\end{lemma}
We have the following result.
\begin{proposition} Let $X$ be a  symmetric random variable with mean $\mu$ and
	variance $\sigma^2$. Suppose GlueVaR has the distortion function $\mathcal{K}_{\beta,\alpha}^{h_1,h_2}$ defined by Eq. (2.1). Then, the following statements hold:\\
	(i) When \( k_2\geq k_1\) and \( h_2\geq 1-\alpha \) or $\max\{k_2,k_3\}<k_1$ and $k_2\geq1$, we have
	\begin{eqnarray*}
		\inf_{X\in V_S(\mu,\sigma)}\rho_{\mathcal{K}_{\beta,\alpha}^{h_1,h_2}}[X]=\mu,
	\end{eqnarray*}	
	and the best-case rv $X^*= \arg\inf_{X\in V_S(\mu,\sigma)}\rho_{\mathcal{K}_{\beta,\alpha}^{h_1,h_2}}[X] $  is  with
	\begin{eqnarray}
		F_{X^*}^{-1+}(p)=\mu, p\in[0,1].
	\end{eqnarray}	
	(ii) When \( k_2\geq k_1\) and \( h_2<1-\alpha \), we have\\
	For $\frac{1}{2}\le\alpha<\beta< 1 $
	\begin{eqnarray*}
		\inf_{X\in V_S(\mu,\sigma)}\rho_{\mathcal{K}_{\beta,\alpha}^{h_1,h_2}}[X]=\mu+\sigma \frac{h_2-1+\alpha}{\alpha\sqrt{2(1-\alpha) }},
	\end{eqnarray*}	
	and the best-case rv $X^*= \arg\inf_{X\in V_S(\mu,\sigma)}\rho_{\mathcal{K}_{\beta,\alpha}^{h_1,h_2}}[X] $  is  with
	\begin{eqnarray}
		F_{X^*}^{-1+}(p)=\left\{\begin{array}{ll}
			\mu+\sigma \sqrt{\frac{1}{2(1-\alpha)} },  \ &{\rm if}\,\, p\in \left[0,1-\alpha\right),\\
			\mu,  \ &{\rm if}\,\, p\in [1-\alpha,\alpha),\\
			\mu-\sigma \sqrt{\frac{1}{2(1-\alpha)} }, \ &{\rm if}\, \,  p\in [\alpha,1].\\
		\end{array}
		\right.
	\end{eqnarray}
	For $0<\alpha<\beta<\frac{1}{2}$
	\begin{eqnarray*}
		\inf_{X\in V_S(\mu,\sigma)}\rho_{\mathcal{K}_{\beta,\alpha}^{h_1,h_2}}[X]=\mu+\sigma \frac{h_2+\alpha-1}{(1-\alpha)\sqrt{2\alpha }},
	\end{eqnarray*}
	and the best-case rv $X^*= \arg\inf_{X\in V_S(\mu,\sigma)}\rho_{\mathcal{K}_{\beta,\alpha}^{h_1,h_2}}[X] $  is  with
	\begin{eqnarray}
		F_{X^*}^{-1+}(p)=\left\{\begin{array}{ll}
			\mu+\sigma \sqrt{\frac{1}{2\alpha} },  \ &{\rm if}\,\, p\in \left[0,\alpha\right),\\
			\mu,  \ &{\rm if}\,\, p\in \left[\alpha,1-\alpha\right),\\
			\mu-\sigma \sqrt{\frac{1}{2\alpha} }, \ &{\rm if}\, \,  p\in [1-\alpha,1].\\
		\end{array}
		\right.
	\end{eqnarray}
	(iii) When \( k_3>k_1>k_2 \), we have\\
	For $\frac{1}{2}\le\alpha<\beta< 1$
	\begin{eqnarray*}
		\inf_{X\in V_S(\mu,\sigma)}\rho_{\mathcal{K}_{\beta,\alpha}^{h_1,h_2}}[X]=\mu-\sigma\sqrt{\frac{\omega}{2\alpha^2}},
	\end{eqnarray*}
	and the best-case rv $X^*= \arg\inf_{X\in V_S(\mu,\sigma)}\rho_{\mathcal{K}_{\beta,\alpha}^{h_1,h_2}}[X] $  is  with
	\begin{eqnarray}
		F_{X^*}^{-1+}(p)=\left\{\begin{array}{ll}
			\mu-\sigma\frac{\alpha\beta+(h_1-1)\alpha+(h_2-1)(1-\beta)}{(1-\beta)\sqrt{2\omega}} ,  \ &{\rm if}\,\, p\in \left[0,1-\beta\right),\\
			\mu-\sigma \frac{(1-h_1)\alpha(1+\alpha-\beta)-(1-h_2)\beta}{(\beta-\alpha)\sqrt{2\omega}} ,  \ &{\rm if}\,\, p\in \left[1-\beta,1-\alpha\right),\\
			\mu,  \ &{\rm if}\,\, p\in \left[1-\alpha,\alpha\right),\\
			\mu+\sigma \frac{(1-h_1)\alpha(1+\alpha-\beta)-(1-h_2)\beta}{(\beta-\alpha)\sqrt{2\omega}} ,  \ &{\rm if}\,\, p\in \left[\alpha,\beta\right),\\
			\mu+\sigma\frac{\alpha\beta+(h_1-1)\alpha+(h_2-1)(1-\beta)}{(1-\beta)\sqrt{2\omega}} ,  \ &{\rm if}\,\, p\in \left[\beta,1\right],\\
		\end{array}
		\right.
	\end{eqnarray}
	where $\omega=\frac{[\alpha\beta+(h_1-1)\alpha+(h_2-1)(1-\beta)]^2}{1-\beta}+\frac{[(1-h_1)\alpha(1+\alpha-\beta)-(1-h_2)\beta]^2}{\beta-\alpha}$.\\
	For $0<\alpha<\beta<\frac{1}{2}$
	\begin{eqnarray*}
		\inf_{X\in V_S(\mu,\sigma)}\rho_{\mathcal{K}_{\beta,\alpha}^{h_1,h_2}}[X]=\mu-\sigma\sqrt{\frac{\nu}{2(1-\beta)^2}},
	\end{eqnarray*}
	and the best-case rv $X^*= \arg\inf_{X\in V_S(\mu,\sigma)}\rho_{\mathcal{K}_{\beta,\alpha}^{h_1,h_2}}[X] $  is  with
	\begin{eqnarray}
		F_{X^*}^{-1+}(p)=\left\{\begin{array}{ll}
			\mu-\sigma\frac{\alpha\beta+(h_1-1)\alpha+(h_2-1)(1-\beta)}{\alpha\sqrt{2\nu}} ,  \ &{\rm if}\,\, p\in \left[0,\alpha\right),\\
			\mu-\sigma \frac{(1-h_1)\alpha(1+\alpha-\beta)-(1-h_2)\beta}{(\beta-\alpha)\sqrt{2\nu}} ,  \ &{\rm if}\,\, p\in \left[\alpha,\beta\right),\\
			\mu,  \ &{\rm if}\,\, p\in \left[\beta,1-\beta\right),\\
			\mu+\sigma \frac{(1-h_1)\alpha(1+\alpha-\beta)-(1-h_2)\beta}{(\beta-\alpha)\sqrt{2\nu}} ,  \ &{\rm if}\,\, p\in \left[1-\beta,1-\alpha\right),\\
			\mu+\sigma\frac{\alpha\beta+(h_1-1)\alpha+(h_2-1)(1-\beta)}{\alpha\sqrt{2\nu}} ,  \ &{\rm if}\,\, p\in \left[1-\alpha,1\right],\\
		\end{array}
		\right.
	\end{eqnarray}
	where $\nu=\frac{[\alpha\beta+(h_1-1)\alpha+(h_2-1)(1-\beta)]^2}{\alpha}+\frac{[(1-h_1)\alpha(1+\alpha-\beta)-(1-h_2)\beta]^2}{\beta-\alpha}$.\\
	(iv) When \( k_3=k_1>k_2 \) or $\max\{k_2,k_3\}<k_1$ and $k_2<1$, we have\\
	For $\frac{1}{2}\le\alpha<\beta< 1$\\
	\begin{eqnarray*}
		\inf_{X\in V_S(\mu,\sigma)}\rho_{\mathcal{K}_{\beta,\alpha}^{h_1,h_2}}[X]=\mu+\sigma \frac{h_1-1+\beta}{\beta\sqrt{2(1-\beta) }},
	\end{eqnarray*}
	and the best-case rv $X^*= \arg\inf_{X\in V_S(\mu,\sigma)}\rho_{\mathcal{K}_{\beta,\alpha}^{h_1,h_2}}[X] $  is  with
	\begin{eqnarray}
		F_{X^*}^{-1+}(p)=\left\{\begin{array}{ll}
			\mu+\sigma \sqrt{\frac{1}{2(1-\beta)} },  \ &{\rm if}\,\, p\in \left[0,1-\beta\right),\\
			\mu,  \ &{\rm if}\,\, p\in \left[1-\beta,\beta\right),\\
			\mu-\sigma \sqrt{\frac{1}{2(1-\beta)} }, \ &{\rm if}\, \,  p\in [\beta,1].\\
		\end{array}
		\right.
	\end{eqnarray}	
	For $0<\alpha<\beta<\frac{1}{2}$
	\begin{eqnarray*}
		\inf_{X\in V_S(\mu,\sigma)}\rho_{\mathcal{K}_{\beta,\alpha}^{h_1,h_2}}[X]=\mu+\sigma \frac{h_1-1+\beta}{(1-\beta)\sqrt{2\beta}},
	\end{eqnarray*}	
	and the best-case rv $X^*= \arg\inf_{X\in V_S(\mu,\sigma)}\rho_{\mathcal{K}_{\beta,\alpha}^{h_1,h_2}}[X] $  is  with
	\begin{eqnarray}
		F_{X^*}^{-1+}(p)=\left\{\begin{array}{ll}
			\mu+\sigma \sqrt{\frac{1}{2\beta} },  \ &{\rm if}\,\, p\in \left[0,\beta\right),\\
			\mu,  \ &{\rm if}\,\, p\in \left[\beta,1-\beta\right),\\
			\mu-\sigma \sqrt{\frac{1}{2\beta} }, \ &{\rm if}\, \,  p\in [1-\beta,1].\\
		\end{array}
		\right.
	\end{eqnarray}
\end{proposition}
{\bf Proof}.
Using the same arguments as Proposition 4.1, we consider the following classification of cases consistent with Proposition 4.1:\\
{\bf (i)} \( k_2\geq k_1\) and \( h_2\geq 1-\alpha \) or $\max\{k_2,k_3\}<k_1$ and $k_2\geq1$.\\
In this case,
\[ 	{\mathcal{K}_{\beta,\alpha}^{h_1,h_2}}_*^{'}(p)=1, p\in[0,1] \]
and
\[ 	{\mathcal{K}_{\beta,\alpha}^{h_1,h_2}}_*^{'}(1-p)=1, p\in[0,1]. \]
Using Lemma 4.2, we have
\[ \int_0^1\left({\mathcal{K}_{\beta,\alpha}^{h_1,h_2}}_*^{'}(p)-{\mathcal{K}_{\beta,\alpha}^{h_1,h_2}}_*^{'}(1-p)\right)^2dp=0. \]
So
\[ 	\inf_{X\in V_S(\mu,\sigma)}\rho_{\mathcal{K}_{\beta,\alpha}^{h_1,h_2}}[X]=\mu,  \]
and the best-case rv $X^*= \arg\inf_{X\in V_S(\mu,\sigma)}\rho_{\mathcal{K}_{\beta,\alpha}^{h_1,h_2}}[X] $  is  with (4.7).\\
{\bf (ii)} $k_2\geq k_1$ and $h_2< 1-\alpha$.\\
In this case,
\begin{eqnarray*}
	{\mathcal{K}_{\beta,\alpha}^{h_1,h_2}}_*^{'}(p)=\left\{\begin{array}{ll}
		\frac{h_2}{1-\alpha}, \ &{\rm if}\, \,  p\in [0,1-\alpha),\\
		\frac{1-h_2}{\alpha}, \ &{\rm if}\, \,  p\in [1-\alpha,1],\\
	\end{array}
	\right.
\end{eqnarray*}	
and
\begin{eqnarray*}
	{\mathcal{K}_{\beta,\alpha}^{h_1,h_2}}_*^{'}(1-p)=\left\{\begin{array}{ll}
		\frac{1-h_2}{\alpha}, \ &{\rm if}\, \,  p\in [0,\alpha],\\
		\frac{h_2}{1-\alpha}, \ &{\rm if}\, \,  p\in(\alpha,1].\\
	\end{array}
	\right.
\end{eqnarray*}	
Using Lemma 4.2, for $\frac{1}{2}\le\alpha<\beta< 1 $, we have
\[ \int_0^1\left({\mathcal{K}_{\beta,\alpha}^{h_1,h_2}}_*^{'}(p)-{\mathcal{K}_{\beta,\alpha}^{h_1,h_2}}_*^{'}(1-p)\right)^2dp=\frac{2(h_2-1+\alpha)^2}{\alpha^2(1-\alpha)}. \]
So
\[\inf_{X\in V_S(\mu,\sigma)}\rho_{\mathcal{K}_{\beta,\alpha}^{h_1,h_2}}[X]=\mu+\sigma \frac{h_2-1+\alpha}{\alpha\sqrt{2(1-\alpha) }}, \]
and the best-case rv $X^*= \arg\inf_{X\in V_S(\mu,\sigma)}\rho_{\mathcal{K}_{\beta,\alpha}^{h_1,h_2}}[X] $  is  with (4.8).\\
Similarly, for $0<\alpha<\beta<\frac{1}{2}$,
\[ \int_0^1\left({\mathcal{K}_{\beta,\alpha}^{h_1,h_2}}_*^{'}(p)-{\mathcal{K}_{\beta,\alpha}^{h_1,h_2}}_*^{'}(1-p)\right)^2dp=\frac{2[\alpha h_2-(1-\alpha)(1-h_2)]^2}{\alpha(1-\alpha)^2}. \]
So
\[\inf_{X\in V_S(\mu,\sigma)}\rho_{\mathcal{K}_{\beta,\alpha}^{h_1,h_2}}[X]=\mu+\sigma \frac{h_2+\alpha-1}{(1-\alpha)\sqrt{2\alpha }}, \]
and the best-case rv $X^*= \arg\inf_{X\in V_S(\mu,\sigma)}\rho_{\mathcal{K}_{\beta,\alpha}^{h_1,h_2}}[X] $  is  with (4.9).\\
{\bf (iii)} $k_2< k_1<k_3$.\\
In this case,
\begin{eqnarray*}
	{\mathcal{K}_{\beta,\alpha}^{h_1,h_2}}_*^{'}(p)=\left\{\begin{array}{ll}
		\frac{h_1}{1-\beta}, \ &{\rm if}\, \,  p\in [0,1-\beta),\\
		h_1+\frac{h_2-h_1}{\beta-\alpha}, \ &{\rm if}\, \,  p\in [1-\beta,1-\alpha),\\
		\frac{1-h_2}{\alpha}+1, \ &{\rm if}\, \,  p\in [1-\alpha,1].\\
	\end{array}
	\right.
\end{eqnarray*}	
and
\begin{eqnarray*}
	{\mathcal{K}_{\beta,\alpha}^{h_1,h_2}}_*^{'}(1-p)=\left\{\begin{array}{ll}
		\frac{1-h_2}{\alpha}+1, \ &{\rm if}\, \,  p\in [0,\alpha],\\
		h_1+\frac{h_2-h_1}{\beta-\alpha}, \ &{\rm if}\, \,  p\in (\alpha,\beta],\\
		\frac{h_1}{1-\beta}, \ &{\rm if}\, \,  p\in (\beta,1].\\
	\end{array}
	\right.
\end{eqnarray*}	
Using Lemma 4.2, for $\frac{1}{2}\le\alpha<\beta< 1 $, we have
\begin{eqnarray*}
	\int_0^1\left({\mathcal{K}_{\beta,\alpha}^{h_1,h_2}}_*^{'}(p)-{\mathcal{K}_{\beta,\alpha}^{h_1,h_2}}_*^{'}(1-p)\right)^2dp=\frac{2\omega}{\alpha^2},
\end{eqnarray*}	
where $\omega=\frac{[\alpha\beta+(h_1-1)\alpha+(h_2-1)(1-\beta)]^2}{1-\beta}+\frac{[(1-h_1)\alpha(1+\alpha-\beta)-(1-h_2)\beta]^2}{\beta-\alpha}$.\\
So
\[\inf_{X\in V_S(\mu,\sigma)}\rho_{\mathcal{K}_{\beta,\alpha}^{h_1,h_2}}[X]=\mu-\sigma\sqrt{\frac{\omega}{2\alpha^2}}, \]
and the best-case rv $X^*= \arg\inf_{X\in V_S(\mu,\sigma)}\rho_{\mathcal{K}_{\beta,\alpha}^{h_1,h_2}}[X] $  is  with (4.10).\\
Similarly, for $0<\alpha<\beta<\frac{1}{2}$,
\begin{eqnarray*}
	\int_0^1\left({\mathcal{K}_{\beta,\alpha}^{h_1,h_2}}_*^{'}(p)-{\mathcal{K}_{\beta,\alpha}^{h_1,h_2}}_*^{'}(1-p)\right)^2dp=\frac{2\nu}{(1-\beta)^2},
\end{eqnarray*}	
where $\nu=\frac{[\alpha\beta+(h_1-1)\alpha+(h_2-1)(1-\beta)]^2}{\alpha}+\frac{[(1-h_1)\alpha(1+\alpha-\beta)-(1-h_2)\beta]^2}{\beta-\alpha}$.\\
So
\[\inf_{X\in V_S(\mu,\sigma)}\rho_{\mathcal{K}_{\beta,\alpha}^{h_1,h_2}}[X]=\mu-\sigma\sqrt{\frac{\nu}{2(1-\beta)^2}}, \]
and the best-case rv $X^*= \arg\inf_{X\in V_S(\mu,\sigma)}\rho_{\mathcal{K}_{\beta,\alpha}^{h_1,h_2}}[X] $  is  with (4.11).\\
{\bf (iv)} \( k_3=k_1>k_2 \) or $\max\{k_2,k_3\}<k_1$ and $k_2<1$.\\
In this case,
\begin{eqnarray*}
	{\mathcal{K}_{\beta,\alpha}^{h_1,h_2}}_*^{'}(p)=\left\{\begin{array}{ll}
		\frac{h_1}{1-\beta}, \ &{\rm if}\, \,  p\in [0,1-\beta),\\
		\frac{1-h_1}{\beta}, \ &{\rm if}\, \,  p\in [1-\beta,1].\\
	\end{array}
	\right.
\end{eqnarray*}	
and
\begin{eqnarray*}
	{\mathcal{K}_{\beta,\alpha}^{h_1,h_2}}_*^{'}(1-p)=\left\{\begin{array}{ll}
		\frac{1-h_1}{\beta}, \ &{\rm if}\, \,  p\in [0,\beta],\\
		\frac{h_1}{1-\beta}, \ &{\rm if}\, \,  p\in (\beta,1].\\
	\end{array}
	\right.
\end{eqnarray*}	
Using Lemma 4.2, for $\frac{1}{2}\le\alpha<\beta< 1 $, we have
\[ \int_0^1\left({\mathcal{K}_{\beta,\alpha}^{h_1,h_2}}_*^{'}(p)-{\mathcal{K}_{\beta,\alpha}^{h_1,h_2}}_*^{'}(1-p)\right)^2dp=\frac{2(h_1-1+\beta)^2}{\beta^2(1-\beta)}. \]
So
\[\inf_{X\in V_S(\mu,\sigma)}\rho_{\mathcal{K}_{\beta,\alpha}^{h_1,h_2}}[X]=\mu+\sigma \frac{h_1-1+\beta}{\beta\sqrt{2(1-\beta) }}, \]
and the best-case rv $X^*= \arg\inf_{X\in V_S(\mu,\sigma)}\rho_{\mathcal{K}_{\beta,\alpha}^{h_1,h_2}}[X] $  is  with (4.12).\\
Similarly, for $0<\alpha<\beta<\frac{1}{2}$
\[ \int_0^1\left({\mathcal{K}_{\beta,\alpha}^{h_1,h_2}}_*^{'}(p)-{\mathcal{K}_{\beta,\alpha}^{h_1,h_2}}_*^{'}(1-p)\right)^2dp=\frac{2(h_1-1+\beta)^2}{\beta(1-\beta)^2}. \]
So
\[\inf_{X\in V_S(\mu,\sigma)}\rho_{\mathcal{K}_{\beta,\alpha}^{h_1,h_2}}[X]=\mu+\sigma \frac{h_1-1+\beta}{(1-\beta)\sqrt{2\beta}}, \]
and the best-case rv $X^*= \arg\inf_{X\in V_S(\mu,\sigma)}\rho_{\mathcal{K}_{\beta,\alpha}^{h_1,h_2}}[X] $  is  with (4.13). This completes the proof of Proposition 4.2.
\begin{remark}
Applying Proposition 4.2, we obtain the best-case VaR, TVaR, and RVaR for symmetric distributions as follows.\\
(i) When $0<\alpha<\beta<\frac{1}{2}$, we have
\begin{eqnarray}
	\inf_{X\in V_S(\mu,\sigma)}RVaR_{\alpha,\beta}[X]=\inf_{X\in V_S(\mu,\sigma)}\rho_{\mathcal{K}_{\beta,\alpha}^{0,1}}[X]=\mu-\sigma \sqrt{\frac{1}{2\beta}},
\end{eqnarray}
and the best-case rv $X^*= \arg\inf_{X\in V_S(\mu,\sigma)}\rho_{\mathcal{K}_{\beta,\alpha}^{0,1}}[X] $ is  with
\begin{eqnarray}
	F_{X_*}^{-1+}(p)=\left\{\begin{array}{ll}
		\mu+\sigma \sqrt{\frac{1}{2\beta} },  \ &{\rm if}\,\, p\in \left[0,\beta\right),\\
		\mu,  \ &{\rm if}\,\, p\in \left[\beta,1-\beta\right),\\
		\mu-\sigma \sqrt{\frac{1}{2\beta} }, \ &{\rm if}\, \,  p\in [1-\beta,1].\\
	\end{array}
	\right.
\end{eqnarray}

In particular, letting $\beta\to \alpha$ in (4.14) and (4.15), respectively, for $0<\alpha<\frac{1}{2}$, we obtain
\[ 	\inf_{X\in V_S(\mu,\sigma)}VaR_{\alpha}[X]=\inf_{X\in V_S(\mu,\sigma)}\rho_{\mathcal{K}_{\alpha,\alpha}^{0,0}}[X]=\mu-\sigma\sqrt{ \frac{1}{2\alpha }},\]
and the best-case rv $X^*= \arg\inf_{X\in V_S(\mu,\sigma)}\rho_{\mathcal{K}_{\alpha,\alpha}^{0,0}}[X] $  is  with
\begin{eqnarray*}
	F_{X^*}^{-1+}(p)=\left\{\begin{array}{ll}
		\mu+\sigma \sqrt{\frac{1}{2\alpha} },  \ &{\rm if}\,\, p\in \left[0,\alpha\right),\\
		\mu,  \ &{\rm if}\,\, p\in \left[\alpha,1-\alpha\right),\\
		\mu-\sigma \sqrt{\frac{1}{2\alpha} }, \ &{\rm if}\, \,  p\in [1-\alpha,1].\\
	\end{array}
	\right.
\end{eqnarray*}
(ii) When $\frac{1}{2}\le\alpha<\beta< 1 $, we have
\begin{eqnarray}
	\inf_{X\in V_S(\mu,\sigma)}RVaR_{\alpha,\beta}[X]=\inf_{X\in V_S(\mu,\sigma)}\rho_{\mathcal{K}_{\beta,\alpha}^{0,1}}[X]=\mu-\sigma
	\sqrt{\frac{1-\beta}{2\beta^2 }},
\end{eqnarray}
and the best-case rv $X^*= \arg\inf_{X\in V_S(\mu,\sigma)}\rho_{\mathcal{K}_{\beta,\alpha}^{0,1}}[X] $  is  with
\begin{eqnarray}
	F_{X_*}^{-1+}(p)=\left\{\begin{array}{ll}
		\mu+\sigma \sqrt{\frac{1}{2(1-\beta)} },  \ &{\rm if}\,\, p\in \left[0,1-\beta\right),\\
		\mu,  \ &{\rm if}\,\, p\in \left[1-\beta,\beta\right),\\
		\mu-\sigma \sqrt{\frac{1}{2(1-\beta)} }, \ &{\rm if}\, \,  p\in [\beta,1].\\
	\end{array}
	\right.
\end{eqnarray}

In particular, letting $\beta\to \alpha$ in (4.16) and (4.17), respectively, for $\frac{1}{2}\le\alpha<\beta< 1 $, we obtain
\[ 	\inf_{X\in V_S(\mu,\sigma)}VaR_{\alpha}[X]=\inf_{X\in V_S(\mu,\sigma)}\rho_{\mathcal{K}_{\alpha,\alpha}^{0,0}}[X]=\mu-\sigma \sqrt{\frac{1-\alpha}{2\alpha^2 }},\]
and the best-case rv $X^*= \arg\inf_{X\in V_S(\mu,\sigma)}\rho_{\mathcal{K}_{\alpha,\alpha}^{0,0}}[X] $  is  with
\begin{eqnarray*}
	F_{X^*}^{-1+}(p)=\left\{\begin{array}{ll}
		\mu+\sigma \sqrt{\frac{1}{2(1-\alpha)} },  \ &{\rm if}\,\, p\in \left[0,1-\alpha\right),\\
		\mu,  \ &{\rm if}\,\, p\in \left[1-\alpha,\alpha\right),\\
		\mu-\sigma \sqrt{\frac{1}{2(1-\alpha)} }, \ &{\rm if}\, \,  p\in [\alpha,1].\\
	\end{array}
	\right.
\end{eqnarray*}

Additionally, letting $\beta\to 1$ in (4.16) and (4.17), respectively, we get
\[ 	\inf_{X\in V_S(\mu,\sigma)}TVaR_{\alpha}[X]=\inf_{X\in V_S(\mu,\sigma)}\rho_{\mathcal{K}_{\alpha,\alpha}^{1,1}}[X]=\mu,   \]
and the best-case rv $X^*= \arg\inf_{X\in V_S(\mu,\sigma)}\rho_{\mathcal{K}_{\alpha,\alpha}^{1,1}}[X]$ is  with
\[ 	F_{X^*}^{-1+}(p)=\mu,p\in[0,1]. \]

The above results are consistent with the results of Zhao et al. (2024), Li et al. (2018) and so on for symmetric distributions.
\end{remark}
 \section {Conclusions and Future Work }\label{intro}

	In this paper, we obtain the closed-form solutions for the best- and worst-case GlueVaR distortion risk measure and the corresponding extreme-case distributions with the knowledge of the first two moments of the underlying distributions. Especially, the extreme-case distributions can be characterized by the envelopes of the corresponding distortion functions. Even with shape constraints, such as symmetry of the underlying distributions, we can still derive the extremal cases estimates of GlueVaR distortion risk measure when the first two moments are known. In addition,  we further apply our results to obtain explicit bounds on VaR, TVaR and RVaR. Our findings broaden the applicability of closed-form solutions for extremal cases of distortion risk measures, and significantly improve existing moment bounds of the GlueVaR distortion risk measure under both moment and shape constraints. We plan to study explicit bounds for GlueVaR distortion riskmetrics, whose distortion function is a bounded variation function on [0,1] (see, e.g.,  Wang et al.  (2020)). In addition, Zhao et al. (2024) not only derive extremal cases of general class of distortion risk measures for general, symmetric random variables but also for unimodal, unimodal-symmetric random variables. Inspired by this, we consider the computation of extreme cases of GlueVaR distortion risk measure for unimodal and unimodal-symmetric random variables. We are currently investigating these interesting questions and hope to report our findings in future papers.

 \noindent{\bf CRediT authorship contribution statement}

 {\bf Mengshuo Zhao:} Investigation, Methodology,  Writing--original draft.
 {\bf Chuancun Yin:} Conceptualization, Investigation, Methodology,  Validation, Writing--original draft, Writing--review \& editing.

 \noindent {\bf Declaration of competing interest}

 There is no competing interest.

 \noindent {\bf Data availability}

 No data was used for the research described in the article.

 \noindent{\bf Acknowledgements}\,  
 This research was supported by the National Natural Science Foundation of China (No. 12071251, 12401616).

\bibliographystyle{model1-num-names}

\end{document}